\documentclass[%
 reprint,
superscriptaddress,
amsmath,amssymb,
aps,
pra,
]{revtex4-1}
\usepackage{graphicx}
\usepackage{dcolumn}
\usepackage{bm}
\usepackage{float}
\usepackage{multirow}

\usepackage[dvipdfm,colorlinks,linkcolor=blue, urlcolor=blue, anchorcolor=blue, citecolor=blue]{hyperref}

\begin{document}


\title{Noise-induced quantum synchronization of spin chain with periodic boundary}
\author{Zhan-Ting Zhang}%
\affiliation{College of Physics and Electronic Engineering, Northwest Normal University, Lanzhou, 730070, China}
\author{Ying-Bo Gao}%
\affiliation{College of Physics and Electronic Engineering, Northwest Normal University, Lanzhou, 730070, China}

\author{Fu-Quan Dou}
\email{doufq@nwnu.edu.cn}
\affiliation{College of Physics and Electronic Engineering, Northwest Normal University, Lanzhou, 730070, China}
\affiliation{Gansu Provincial Research Center for Basic Disciplines of Quantum Physics, Lanzhou, 730070, China}

\begin{abstract}
Quantum synchronization offers new possibilities for the exploration of collective dynamics in many-body systems. However, achieving synchronization in many-body quantum systems still faces numerous challenges. Here, we focus on the synchronization behavior of quantum spin chain with periodic boundary conditions under the influence of local Gaussian white noise. The necessary conditions for synchronization of local spin observables when noise acts on individual spin and two spins are obtained. The degree of synchronization between the expectation values of the local spin observables is characterized by utilizing the Pearson correlation coefficient, while the frequency of oscillation is determined through the application of fast Fourier transformation. We also discuss qualitatively the effect of system parameters on synchronization time. Despite the presence of noise leading to decoherence in the system, entanglement between synchronous and antisynchronous spins still persists. Our results provide valuable insights toward the realization of quantum synchronization in many-body quantum systems.
\end{abstract}
\maketitle
\section{Introduction}
Synchronization is a collective behavior that emerges from two or more dynamically coupled units \cite{WU20241}, which may emerge spontaneously between systems, induced by interactions, dissipation \cite{PhysRevA.94.052118,PhysRevA.88.042115,PhysRevLett.131.190402,PhysRevLett.121.063601}, or driven by an external field acting as a pacemaker that tries to impose its rhythm on the system \cite{PhysRevLett.112.094102,PhysRevA.109.023502,PhysRevResearch.5.033209,PhysRevA.101.062104}. It was first recorded in the $17th$ century by Huygens, who observed consistent oscillations of two pendulums with different swing frequencies after suspended them on the same beam \cite{huygens1916oeuvres}. Since then, synchronization phenomena have been observed in various fields such as physics, biology, engineering and medicine \cite{pikovsky2001universal,na2000sound,PhysRevLett.81.3291,10.1063/1.2956986}.

During the past few years, synchronization research has delved into the quantum realm which has been studied in diverse systems, encompassing systems with a classical analog, such as the Kuramoto model \cite{RevModPhys.77.137,PhysRevE.90.052904,PhysRevE.93.062315,PhysRevLett.93.084102}, Van der Pol oscillators \cite{PhysRevLett.111.234101,PhysRevResearch.5.023021,PhysRevResearch.1.033012,PhysRevResearch.5.023021} as well as inherently quantum mechanical systems, like spin chains \cite{PhysRevLett.132.010402,PhysRevLett.129.250601,PhysRevLett.132.196601,PhysRevA.88.042115} and atomic ensembles \cite{PhysRevA.101.042121,PhysRevLett.123.023604}. In the areas of quantum information \cite{pljonkin2019vulnerability,PhysRevApplied.13.054041,PhysRevA.104.012410,PhysRevB.104.045420}, precision measurement \cite{PhysRevLett.114.103601,PhysRevA.94.052121} and quantum thermal machines \cite{PhysRevE.101.020201,PhysRevA.108.012205}, quantum synchronization is anticipated to have extensive application prospects. Quantum synchronization has been extensively studied, ranging from searching for quantum systems that display synchronization phenomena \cite{PhysRevLett.123.023604,PhysRevA.100.012133}, quantifying the degree of quantum synchronization \cite{PhysRevE.96.012211,Galve2017,PhysRevResearch.2.043287}, to exploring associated quantum technological applications \cite{PhysRevA.111.012410,PhysRevLett.114.103601}.

In addition, the study on quantum synchronization in many-body quantum systems has also become a pivotal research frontier, including explorations of quantum synchronization in two finite spin chains \cite{GHILDIYAL2025130123} and quantum synchronization in multiplex networks \cite{WU20241,ManzanoPaule2018}. As a paradigm of many-body quantum systems, the study of quantum spin chains has a long history that can be traced back to the early explanations of the magnetization behavior of solids. Interesting mathematical structures allow them to exhibit a wealth of properties, including energy spectrum \cite{PhysRevA.89.043608,PhysRevLett.59.259,PhysRevB.6.3444,PhysRevLett.59.259,PhysRevE.80.061109} quantum phase transition \cite{PhysRevLett.112.217204,PhysRevB.102.144437,PhysRevLett.89.127202}, quantum correlations \cite{PhysRevA.102.042206,PhysRevA.98.052303,PhysRevA.82.012106,PhysRevB.83.214416,PhysRevB.107.014207} and quantum transport \cite{RevModPhys.93.025003,PhysRevB.88.205135,PhysRevLett.88.077203}. The unique structure of spin chains has drawn researchers' attention to the synchronization phenomenon within these systems. Investigating quantum synchronization in many-body systems, particularly those with periodic boundary conditions, holds significant potential for enhancing our understanding of the complex dynamics inherent in quantum systems. Despite the growing interest in quantum synchronization, research in this area remains limited and previous studies have mostly focused on few-body spin systems. For example, the spin chains with periodic boundary conditions of lengths three \cite{10.21468/SciPostPhys.12.3.097} and four \cite{PhysRevA.107.032219}. Recently, noise induced synchronization in an isolated many-body quantum system has been investigated by locally applying Gaussian white noise to a quantum spin chain of arbitrary length \cite{PhysRevLett.129.250601}. Experimental observation of noise-induced quantum synchronization in a chain of superconducting transmon qubits with nearest-neighbor interactions has also been achieved \cite{tao2024noise}. Therefore, the following question naturally arises: whether quantum synchronization can be realized in spin chains with periodic boundary conditions.

In this work, we investigate the synchronization phenomena in quantum XX spin chains with periodic boundary conditions subjected to local Gaussian white noise. Stable synchronization or antisynchronization between local spin observables can be achieved when the chain length and the sites of noise application satisfy certain conditions. In that case, local spin observables oscillate with the same frequency, and we perform the fast Fourier transform (FFT) to numerically determine the dominant frequency of the oscillation. The effect of system parameters on synchronization time is examined qualitatively. We also discuss how synchronization influences the dynamics of the system, including Loschmidt echo, purity and trace distance. To analyze the impact of noise on the system's quantum features, we evaluated the mutual information and entanglement between synchronized and antisynchronized spins.

The rest paper is organized as follows. Section \ref{section2} introduces the model. In Sec. \ref{section3} we derive the necessary conditions for synchronization of local spin observables when noise acts on individual and coupled spins, with the Pearson correlation coefficient to characterize the degree of synchronization. The oscillation frequencies of local spin magnetizations are resolved by performing fast Fourier transform. We also qualitatively analyze the impact of system parameters on synchronization time. Then we investigate the time dependence of Loschmidt echo and purity, which are of concern in experimental detection, and explore quantum correlations between synchronous and antisynchronous spins. Finally, a brief summary is given in Sec. \ref{section4}.
\begin{figure}
\centering
\includegraphics[width=0.425\textwidth]{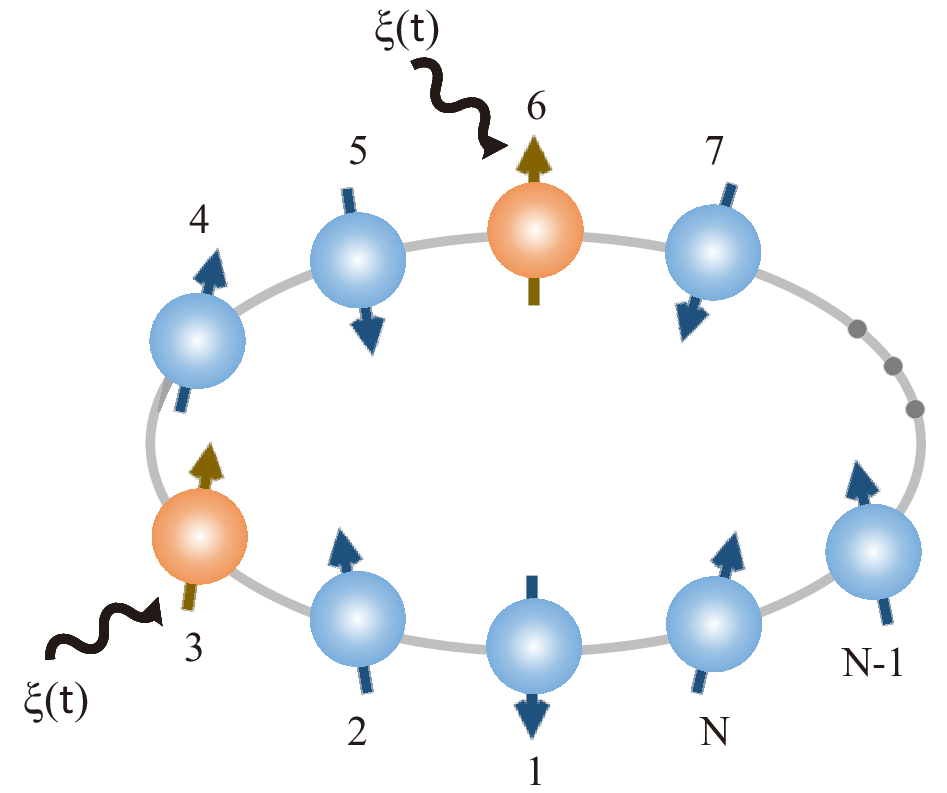}
\caption{Schematic illustration of the investigated spin model. The spin chain features periodic boundary conditions, with each spin interacting with its nearest neighbors through XX interactions. Gaussian white noise $\xi(t)$ are applied locally to the spin chain (orange spheres). The integers $1,2,3,\cdots,N$ denote the sites of the spins. }
\label{fig1}
\end{figure}
\section{Model} \label{section2}
As schematized in Fig. \ref{fig1}, we consider a transverse field quantum XX spin chain with periodic boundary conditions \cite{PhysRevB.97.014301,PhysRevB.70.064409}, subjected to a stochastic perturbations of the form $\xi(t)V$. The Hamiltonian can be written as (hereafter, we set $\hbar=1$)
\begin{eqnarray}
\label{H}
H=H_{0}+\xi(t)V,
\end{eqnarray}
with
\begin{eqnarray}
\label{H0}
H_{0}=-\frac{J}{2}{\sum_{j=1}^{N}}(\sigma_{j}^{x}\sigma_{j+1}^{x}+\sigma_{j}^{y}\sigma_{j+1}^{y})+g\sum_{j=1}^{N}\sigma_{j}^{z}.~
\end{eqnarray}
Here $\xi(t)$ describes noise that couples to the local magnetization spin operator $V$. Similar to Ref. \cite{PhysRevLett.129.250601}, we also choose Gaussian white noise with zero mean and amplitude $\Gamma$ \cite{Levy2020}. The autocorrelation function of the noise is $\langle\xi(t)\xi(t^{'})\rangle=\Gamma\delta(t-t^{'})$. $\sigma_{j}^{\alpha}~(\alpha=x,y,z)$ are Pauli operators, $J$ is the nearest-neighbor interaction between spin-$j$ and spin-$(j+1)$, $g$ represents the strength of the transverse magnetic field and $N$ is the number of spins. The periodic boundary condition is assumed as $\sigma_{N+1}^{\alpha}=\sigma_{1}^{\alpha}$. The quantum spin chain with periodic boundary conditions possesses translational invariance and facilitates extrapolation to the thermodynamic limit \cite{PhysRevB.83.125104}. Various aspects such as entanglement \cite{PhysRevLett.104.095701,PhysRevB.92.054411}, thermodynamic properties \cite{PhysRevB.72.064454,PhysRevB.76.172402} and quantum phase transition \cite{PhysRevLett.96.077206} have been investigated in the system. This many-body system can be implemented using superconducting circuits \cite{tao2024noise} and trapped ions \cite{RevModPhys.93.025001}. In superconducting systems, noise arises from locally tuning the natural frequencies of qubits \cite{PhysRevLett.116.010501}, while in trapped ions, noise is introduced via modulation of the ac-Stark shifts on targeted spin states \cite{PhysRevLett.122.050501}.

To investigate the effects of noise on the quantum spin chain and derive synchronization conditions, we employ the Liouville space formalism to describe the system's time evolution \cite{Gyamfi_2020}. In this formalism, operators such as the density matrix $\rho_{\xi}$ are represented as column vectors denoted by $|\rho_{\xi}\rangle\!\rangle$. These vectorized operators are commonly referred to as supervectors or superkets. The dynamical evolution of the system, described by the von Neumann equation $\dot{\rho}_{\xi}(t)=-i[H_{0}+ \xi(t)V,\rho_{\xi}(t)]$ in Hilbert space, is transformed into the following Schr\"{o}dinger-like equation:
\begin{eqnarray}
\label{density matrix}
|\dot{\rho}_{\xi}(\tau)\rangle\!\rangle=-i[\mathcal{L}_{0}+\xi(\tau)\mathcal{V}]|\rho_{\xi}(\tau)\rangle\!\rangle,
\end{eqnarray}
which takes the form of a Stratonovich stochastic differential equation \cite{KLIMONTOVICH1990515}, with the Liouville superoperator $\mathcal{L}_{0}$ is given by $\mathcal{L}_{0}= [\![H_{0},\mathbb{I}]\!]= H_{0}\otimes\mathbb{I}-\mathbb{I}\otimes H_{0}^{T}$ and the perturbation superoperator by $\mathcal{V}=[\![V,\mathbb{I}]\!]$. The superscript $T$ indicates the transpose operation. $\mathbb{I}$ denotes the identity matrix and the normalized time is given by $\tau=gt$. Upon averaging over an ensemble of noise realizations and converting the Stratonovich stochastic differential equation to an It\^{o} stochastic differential equation \cite{van1981ito}, Eq. (\ref{density matrix}) is rewritten as
\begin{eqnarray}
\label{density matrix average}
|\dot{\rho}(\tau)\rangle\!\rangle=-i(\mathcal{L}_{0}+\frac{\gamma\mathcal{V}^{2}}{2})|\rho(\tau)\rangle\!\rangle,
\end{eqnarray}
where $\rho(\tau)=\langle\rho_{\xi}(\tau)\rangle$ is the averaged density operator and $\gamma=\Gamma/g$ is the reduced noise strength. We treat the effect of Gaussian white noise on the system as a perturbation, with $\gamma$ representing the perturbation intensity.

For unperturbed quantum systems (i.e., $\gamma=0$), $|\rho_{0}(\tau)\rangle\!\rangle=\exp(-i\mathcal{L}_{0}\tau)|\rho_{0}(\tau)\rangle\!\rangle$, according to the spectral decomposition of the free evolution
\begin{eqnarray}
\label{DM spectral decomposition}
|\rho_{0}(\tau)\rangle\!\rangle=\sum\limits_{k,l}e^{-i\Lambda_{kl}\tau}|v_{k},v_{l}\rangle\!\rangle\langle\!\langle v_{k},v_{l}|\rho_{0}(0)\rangle\!\rangle,
\end{eqnarray}
where the eigenfrequencies $\Lambda_{kl}$ and the eigenmodes $|v_{k},v_{l}\rangle\!\rangle$ of the Liouvillian operator $\mathcal{L}_{0}$ are connected to the eigenvalues $\Lambda_{k}$ and the eigenstates $|v_{k}\rangle$ of the Hamiltonian within the Hilbert space via $\Lambda_{kl}=\Lambda_{k}-\Lambda_{l}$ and $|v_{k},v_{l}\rangle\!\rangle=|v_{k}\rangle\otimes|v_{l}\rangle$ \cite{Gyamfi_2020}. $k,l$ are non-negative integers labeling the eigenvalues and eigenstates. It should be noted that eigenfrequencies always come in pairs, satisfying the condition $\Lambda_{kl}=-\Lambda_{lk}$.

For perturbed quantum systems with weak noise ($\gamma\ll1$), which acts as a small perturbation to $\mathcal{L}_{0}$. We are able to determine the first-order eigenmodes and eigenfrequencies of the perturbed system
\begin{eqnarray}
\label{eigenmodes}
\wedge_{kl}^{p}\backsimeq\Lambda_{kl}-i\gamma m_{kl}, |v_{k},v_{l}\rangle\!\rangle^{p}=|v_{k},v_{l}\rangle\!\rangle^{(0)}-
\gamma|v_{k},v_{l}\rangle\!\rangle^{(1)}, \nonumber \\
\end{eqnarray}
where the superscript $p$ labels the perturbed quantities. According to Eq. (\ref{eigenmodes}), eigenfrequencies of this system experience selective exponential decay with rate $\gamma m_{kl}$. Stable synchronization is achieved when only one mode remains, with all others decaying to zero. This results in a decoherence-free subspace that contains only a single eigenmode \cite{PhysRevLett.81.2594,Lidar2003}.
\section{Noise induced quantum synchronization} \label{section3}
\subsection{Stable synchronization condition}
We are mainly concern about the synchronization of the local spin magnetizations $\langle\sigma_{j}^{z}\rangle$ in the system (\ref{H}). In order to examine the influence of noise on the quantum spin chain, we map the interacting spin-$1/2$ systems to a chain of non-interacting spinless fermions and derive the evolution equation of $\langle\sigma_{j}^{z}\rangle$ in Liouville space.

The connection between the abstract Liouvillian eigenstates (in Liouville space) and the physical qubit eigenstates (in Hilbert space) is established by projecting the superoperator $|\rho(\tau)\rangle\!\rangle$ onto the superoperator $|\sigma_{j}^{z}\rangle\!\rangle$. We can obtain the time-dependent of $\sigma_{j}^{z}$ as follows:
\begin{eqnarray}
\label{sigmaz}
\sigma_{j}^{z}(\tau)=\langle\!\langle\sigma_{j}^{z}|\rho(\tau)\rangle\!\rangle=\sum_{kl}c_{kl}e^{-i\tilde{\Lambda}_{kl}\tau}\epsilon_{j,kl},
\end{eqnarray}
for the $j$th qubit, the magnetization eigenmodes can be determined through the projection $\epsilon_{j,kl}=\langle\!\langle\sigma_{j}^{z}|v_{k},v_{l}\rangle\!\rangle$ and the coefficients $c_{kl}=\langle\!\langle v_{k},v_{l}|\rho(0) \rangle\!\rangle$ are influenced by the initial excitations. Thus, the magnetization frequencies $\tilde{\Lambda}_{kl}$ represent a subset of $\{\Lambda_{kl}\}$. Initially, we apply a Jordan-Wigner transformation to reduce the Hilbert space dimensionality \cite{PhysRevLett.86.1082}, converting the Hamiltonian $H_{0}$ (described in Eq. (\ref{H0})) into the fermionic representation:
\begin{eqnarray}
\label{H0Jw}
\begin{split}
H_{JW}=&-J\sum_{j=1}^{N-1}(c_{j}^{\dag}c_{j+1}+c_{j+1}^{\dag}c_{j})+g\sum_{j=1}^{N}(2c_{j}^{\dag}c_{j}-\mathbb{I})  &\\
 &+(-1)^{\hat{\mathcal{N}}}(c_{N}^{\dag}c_{1}+c_{1}^{\dag}c_{N}),
\end{split}
\end{eqnarray}
where fermion number operator $\hat{\mathcal{N}}=\sum_{n=1}^{N}c_{n}^{\dag}c_{n}$, $c_{j}^{\dag}$ and $c_{j}$ are the creation and annihilation operators of fermions on $j$th site. These operators are connected to the standard Pauli operators through the following relations:
\begin{eqnarray}
\label{fermion operators}
c_{j}=\left[\prod_{n=1}^{j-1}(-\sigma_{n}^{z})\right]\sigma_{j}^{-},    c_{j}^{\dag}=\left[\prod_{n=1}^{j-1}(-\sigma_{n}^{z})\right]\sigma_{j}^{+}.
\end{eqnarray}
Hamiltonian $H_{JW}$ commutes with the parity operator $(-1)^{\hat{\mathcal{N}}}$, and it is possible to diagonalize both operators in the identical basis. Parity operator $(-1)^{\hat{\mathcal{N}}}$ has only two eigenvalues, $+1$ and $-1$. Hence, the matrix representation of $H_{JW}$ can be expressed in a block-diagonal form, comprising two large blocks that correspond to each of these eigenvalues,
\begin{eqnarray}
\label{H0JW pmatrix}
H_{JW}=\begin{pmatrix}H_{+}&0\\\\0&H_{-}\end{pmatrix}.
\end{eqnarray}
In the case of $H_{+}$, we substitute $+1$ for $(-1)^{\hat{\mathcal{N}}}$, and for $H_{-}$, we replace $(-1)^{\hat{\mathcal{N}}}$ with $-1$. To obtain the full set of eigenpairs for the Hamiltonian $H_{JW}$, we must select half of the eigenpairs from $H_{+}$ (those with even parity) and the other half from $H_{-}$ (those with odd parity). The eigenpairs of $H_{+}$ with odd parity and those of $H_{-}$ with even parity have no physical significance and are therefore disregarded \cite{PhysRevB.97.014301,PhysRevE.92.042115}.

In the $c_{j}$ basis, the evolution of $Z=\langle(c_{j}^{\dag}c_{k})_{1\leq j, k\leq N}\rangle$, follow a von Neumann-type equation \cite{PhysRevA.26.1209}
\begin{eqnarray}
\label{Z}
\dot{Z}(\tau)=i[\Omega,Z(\tau)],
\end{eqnarray}
where $\Omega$ represents a tridiagonal Toeplitz matrix with certain perturbations in the corners. It features values of $[-J,2h,-J]$ along its diagonals and has additional perturbations of $J$ or $-J$ at the off-diagonal corners located at positions $(N,1)$ and $(1,N)$. The diagonal elements of the matrix $Z(\tau)$ give the time evolution of the system's populations. By applying a simple rescaling, we can derive the magnetizations, $\langle\sigma_{j}^{z}\rangle=2Z_{jj}-1$.

Next, we introduce Gaussian white noise $\xi(\tau)$ to a single spin at site $u$ by setting $V=\sigma_{u}^{z}$. In the Jordan-Wigner representation, the equation of motion then takes the form
\begin{eqnarray}
\label{ZJW}
\dot{Z}_{\xi}(\tau)=i[\Omega+2\xi(\tau)Y,Z_{\xi}(\tau)],
\end{eqnarray}
with $Y=|\hat{e}_{u}\rangle\langle\hat{e}_{u}|$ in the canonical basis. Averaging over the noise in Liouville space, we therefore find
\begin{eqnarray}
\label{Zaverage}
|\dot{Z}(\tau)\rangle\!\rangle=(i\mathcal{Q}-2\gamma\mathcal{Y}^{2})|Z(\tau)\rangle\!\rangle,
\end{eqnarray}
where $\mathcal{Q} = [\![\Omega,\mathbb{I}]\!]$, $\mathcal{Y}=[\![Y,\mathbb{I}]\!]$ and $|Z(\tau)\rangle\!\rangle =\langle|Z_{\xi}(\tau)\rangle\!\rangle\rangle$. Equation (\ref{Zaverage}) takes the form of a Schr\"{o}dinger equation (with Hamiltonian $i\mathcal{Q}$ and perturbation $-2\gamma\mathcal{Y}^{2}$). We can employ standard perturbation theory to calculate the decay rates for small noise amplitudes. Ref. [\citenum{10.1007/978-3-030-77493-6_11}] provides the eigenvalues and eigenvectors of the tridiagonal Toeplitz matrix with certain perturbations in the corners, which allows us to derive the complete set of eigenvalues and eigenvectors of the matrix $\Omega$
\begin{flalign}
\label{omega eigenvalues}
&\tilde{\Lambda}_{k}=2g-2J\cos(\frac{k\pi}{N}),\\
&|\varphi_{k}\rangle=\sqrt{\frac{2}{N}}\left(\sin(\frac{k\pi}{N}),\sin(\frac{2k\pi}{N}),\cdots,\sin(\frac{Nk\pi}{N})\right)^{T},
\end{flalign}
where $1\leq k \leq N$. The eigenfrequencies and eigenmodes of the noise-free system are
\begin{eqnarray}
\label{H0 eigenvalues}
\tilde{\Lambda}_{kl}=\tilde{\Lambda}_{k}-\tilde{\Lambda}_{l}=-2J\left(\cos(\frac{k\pi}{N})-\cos(\frac{l\pi}{N})\right),
\end{eqnarray}
\begin{eqnarray}
\label{H0 eigenvector}
|\varphi_{k}, \varphi_{l}\rangle\!\rangle=|\varphi_{k}\rangle\otimes|\varphi_{l}\rangle
=\frac{2}{N} \left(\sin(\frac{k\pi}{N})\sin(\frac{l\pi}{N}),\right.\nonumber \\
\quad \left. \sin(\frac{2k\pi}{N})\sin(\frac{2l\pi}{N}),\cdots,\sin(\frac{Nk\pi}{N})\sin(\frac{Nl\pi}{N})\right)^{T}.
\end{eqnarray}
Therefore the magnetization frequencies $\tilde{\Lambda}_{kl}$ are a subset of the full set of frequencies $\{\tilde{\Lambda}_{kl}\}\subset\{\Lambda_{kl}\}$ and the magnetization modes $|\varphi_{k}, \varphi_{l}\rangle\!\rangle$ are a subspace of the full eigenmodes $\{|v_{k},v_{l}\rangle\!\rangle\}$. For nondegenerate frequencies, the decay rates $m_{kl}^{u}$ can be determined by calculating the expectation value of the perturbation. However, in cases of degenerate frequencies, it is necessary to compute the eigenvalues and eigenvectors of the perturbation matrix. Considering the symmetry of the system's energy levels under periodic boundary conditions \cite{baeriswyl2012quantumxxchaininterface}, we choose $N\in$ even and calculate the decay rates $m_{kl}^{u}$. Details are provided in Appendix \ref{appendix1}.

For one-site noise, we concretely obtain the $m_{kl}^{u}$ for nondegenerate and degenerate frequencies, respectively
\begin{flalign}
\label{decay-non}
m_{kl}^{u}|_{\text{non}} &=\frac{4}{N}\left(\sin(\frac{uk\pi}{N})^{2}+\sin(\frac{ul\pi}{N})^{2}\right) & \nonumber\\
&-\frac{16}{N^{2}}\left(\sin(\frac{uk\pi}{N})^{2}\sin(\frac{ul\pi}{N})^{2}\right),
\end{flalign}
\begin{flalign}
\label{decay-deg}
m_{kl}^{u}|_{\text{deg}} &=\frac{4}{N}\left(\sin(\frac{uk\pi}{N})^{2}+\sin(\frac{ul\pi}{N})^{2}\right) & \nonumber\\
&-\frac{32}{N^{2}}\left(\sin(\frac{uk\pi}{N})^{2}\sin(\frac{ul\pi}{N})^{2}\right).
\end{flalign}
In order to obtain stable synchronization condition, we attempt to determine a configuration $N, u,k,l$ where only one mode remains and all other modes decay to zero. One require that only a single pair $(k,l)$ fulfills $m_{kl}^{u}=0$. We find that this is achieved when
\begin{eqnarray}
\label{decay-zero}
\sin(\frac{uk\pi}{N})=0 \cap \sin(\frac{ul\pi}{N})=0.
\end{eqnarray}
For a single mode, the only possible configuration is
\begin{eqnarray}
\label{one-site}
\frac{N}{3}\in \mathbb{N},~\frac{u}{3}\in \mathbb{N},~k=\frac{N}{3},\nonumber \\
l=2k, ~(for~N\geq6,~N\in even),
\end{eqnarray}
where $\mathbb{N}$ represents the set of natural numbers.

For two-site noise, $V=\sigma_{u}^{z}+\sigma_{v}^{z}$,
\begin{eqnarray}
\label{two-site}
\frac{N}{3}\in \mathbb{N},~\frac{u}{3}\in \mathbb{N},~\frac{v}{3}\in \mathbb{N},~k=\frac{N}{3}\in \mathbb{N}, \nonumber \\
l=2k,~(for~N \geq 6,~N\in even).
\end{eqnarray}

For muti-site noise case, we can also obtain the stable synchronization condition by using the similar manner.
\begin{figure*}[htbp]
\centering
\includegraphics[width=0.9\textwidth]{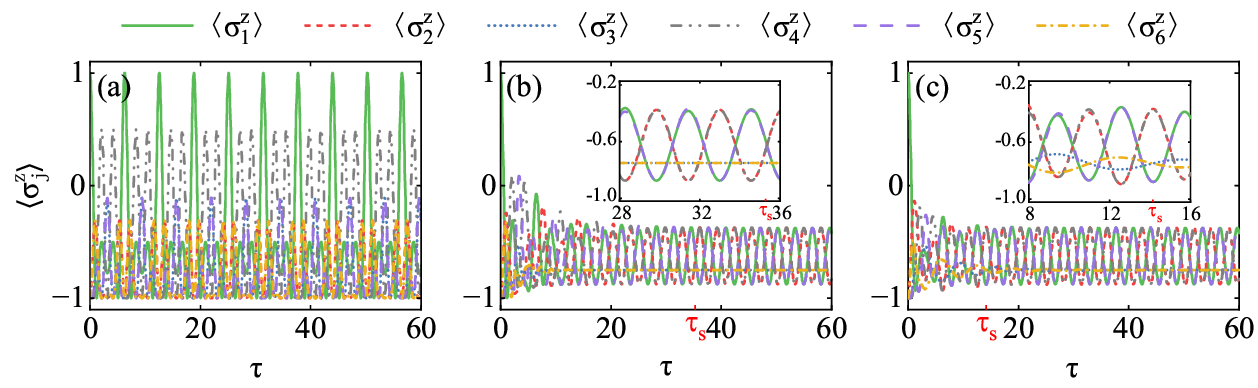}
\caption{The dependence of the local magnetization $\langle\sigma_{i}^{z}\rangle$ on time $\tau$ for three cases: (a) noise-free $\gamma=0$, (b) one-site noise $u=3, \gamma=0.3$ and (c) two-site noise $u=3, v=6, \gamma=0.3$. Each inset illustrates the detailed oscillation of $\sigma_{i}^{z}$ over the period surrounding the synchronization time. The initial state is $|\Psi(0)\rangle=|1\rangle_{1}\otimes\bigotimes|0\rangle_{N-1}$, where $|0\rangle, |1\rangle$ are the ground and excited states of qubit $j$. Other parameters are chosen as $N=6, J=1$ and $h=1$. }
\label{fig2}
\end{figure*}

The time evolution of the magnetization $\langle\sigma_{j}^{z}\rangle$ is shown in Fig. \ref{fig2}. For simplicity, in all our calculations, we take the length of chain $N = 6$. Gaussian white noise acts locally at site $u=3$, as well as at sites $u=3,v=6$, complying with the stable synchronization conditions given in Eq. (\ref{one-site}) and Eq. (\ref{two-site}). The initial state of the system is $|\Psi(0)\rangle=|1\rangle_{1}\otimes\bigotimes|0\rangle_{N-1}$, where $|0\rangle, |1\rangle$ correspond to the ground and excited states of qubit $j$. Figure \ref{fig2}(a) indicates the unperturbed evolution in the absence of noise for comparison. We consider the local magnetization spins to be synchronized when Pearson correlation coefficient $r_{ij}\approx1$ (with an error $10^{-3}$, as detailed in Sec. \ref{c}). After a period of time (the synchronization time $\tau_{s}$), the system will be in the eigenstate denoted as $|v_{k},v_{l}\rangle\rangle^{(s)}$ within the Liouville space and oscillate with the corresponding eigenfrequency $\Lambda_{kl}^{s}$. Before the synchronization time $\tau_{s}$, oscillations are out of phase, and synchronous behavior is not observed. However, for $\tau>\tau_{s}$, stable synchronisation occurs between the local magnetization spin $\langle\sigma_{1}^{z}\rangle$ and $\langle\sigma_{5}^{z}\rangle$, as well as between $\langle\sigma_{2}^{z}\rangle$ and $\langle\sigma_{4}^{z}\rangle$, the magnetization $\langle\sigma_{3}^{z}\rangle$ and $\langle\sigma_{6}^{z}\rangle$ are independent of time in this regime. The insets in Fig. \ref{fig2} show more details, highlighting the evolution of $\sigma_{j}^{z}$ over time near the synchronization time. In addition, the synchronization time $\tau_{s}$ is shorter in the scenario with noise at two sites than in the scenario with noise at one site.

Finally, we explicitly determine the synchronized mode. In the Jordan-Wigner representation, the magnetization eigenmodes $\epsilon_{j,kl} =\langle\!\langle\sigma_{j}^{z}|v_{k},v_{l}\rangle\!\rangle=2\langle\!\langle \hat{e}_{j},\hat{e}_{j}|\varphi_{k},\varphi_{l}\rangle\!\rangle$ can be conveniently calculated by
\begin{eqnarray}
\label{epsilonJW}
|\epsilon_{kl}\rangle =2\sum_{j}\langle\!\langle\hat{e}_{j},\hat{e}_{j}|\varphi_{k},\varphi_{l}\rangle\!\rangle\hat{e}_{j}=
\sum_{j}\varphi_{k}^{j}\varphi_{l}^{j}\hat{e}_{j}, k\neq l, \nonumber \\
\end{eqnarray}
where $\sigma_{j}^{z}(\tau)=2\langle\!\langle\hat{e}_{j},\hat{e}_{j}|Z\rangle\!\rangle-1$. When $k=N/3$ and $l=2N/3$ are inserted into Eq. (\ref{H0 eigenvector}), synchronization occurs, obtaining the $u$-independent non-decaying magnetization mode
\begin{align}
\label{epsilon synchronization}
|\epsilon_{kl}\rangle^{s} &= 2\sum_{j}\langle\!\langle\hat{e}_{j},\hat{e}_{j}|\varphi_{k}^{s},\varphi_{l}^{s}\rangle\!\rangle\hat{e}_{j} \nonumber \\
&= \frac{4}{N}\left(\sin(\frac{\pi}{3})\sin(\frac{2\pi}{3}),\sin(\frac{2\pi}{3})\sin(\frac{4\pi}{3}),\cdots,\right.\nonumber \\&\quad \left.\sin(\frac{N\pi}{3})\sin(\frac{2N\pi}{3})\right)^{T} \nonumber\\
&= \frac{3}{N}(1,-1,0,-1,1,0,\cdots,1,-1,0,-1,1,0)^{T},
\end{align}
where the superscript $s$ labels the synchronized quantities. Magnetization eigenmodes display sixfold periodicity, with two separate groups of local spin magnetizations showing synchronized dynamics. One group consists of local spins at positions $6d+1$ and $6d+5$, while the other group contains local spins at positions $6d+2$ and $6d+4$, where $d$ is an integer in the range $0\leq d\leq(N/6-1)$. These two groups of local spin magnetizations oscillate with a $\pi$ phase difference. For example, when $N=6$, the local spins at sites $1$ and $5$ form one synchronized group, and those at sites $2$ and $4$ form another synchronized group, with these two groups being antisynchronized with each other.

\subsection{Pearson correlation coefficient, Fourier transformation spectrum and synchronization time}\label{c}
Synchronization between a pair of quantum systems emerges via the establishment of coherent oscillations in the expectation values of their local observables \cite{PhysRevA.103.062217}. In fact, one can directly analyze the expectation values of spin observables to determine whether they exhibit synchronized behavior. The Pearson correlation coefficient $r_{ij}$ serves as a valuable tool for quantifying the degree of linear correlation between two time-dependent variables and is commonly used as a metric to characterize synchronization \cite{Galve2017,Benesty2009}. For variables $A_{i}$ and $A_{j}$, the Pearson correlation coefficient is defined as the ratio of their covariance to the product of their respective standard deviations
\begin{eqnarray}
\label{Pearson coefficient}
r_{ij}=\frac{\sum_{t}(A_{i}(t)-\bar{A_{i}})(A_{j}(t)-\bar{A_{j}})}{\sqrt{\sum_{t}(A_{i}(t)-\bar{A_{i}})^2}\sqrt{\sum_{t}(A_{j}(t)-\bar{A_{j}})^2}},
\end{eqnarray}
where $\bar{A_{i}}$ and $\bar{A_{j}}$ are the mean values of $A_{i}$ and $A_{j}$.
\begin{eqnarray}
\label{Ai}
\bar{A_{i}}=\frac{1}{\delta t}\int_{t}^{t+\delta t}A_{i}(t')dt'.
\end{eqnarray}
\begin{figure}
\centering
\includegraphics[width=0.48\textwidth]{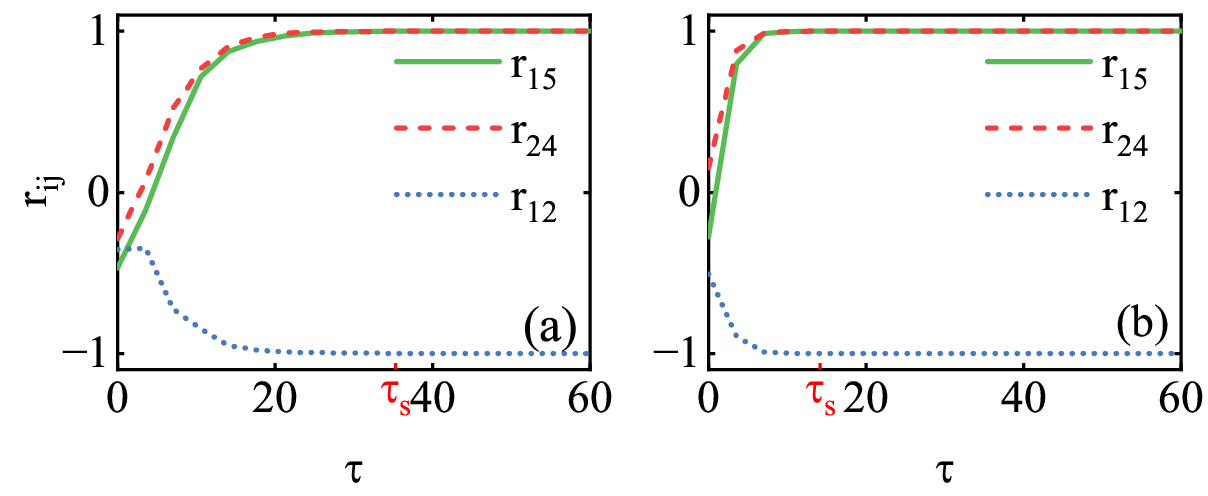}
\caption{Time evolution of the Pearson coefficient $r_{15}$, $r_{24}$ and $r_{12}$ between three different pairs for different cases: (a) one-site noise and (b) two-site noise. The parameters are consistent with those in Fig. \ref{fig2}(b) and \ref{fig2}(c).}
\label{fig3}
\end{figure}
The Pearson coefficient $r_{ij}$ can take a range of values between $1$ and $-1$. A perfect positive correlation ($r_{ij}=1$) indicates that two variables change proportionally in the same direction. A full negative correlation ($r_{ij}=-1$) points out that the variables are anticorrelated, such that an increase in one variable corresponds to a decrease in the other. To further confirme the occurrence of in-phase oscillation between these spins, in Fig \ref{fig3} we illustrate the time evolution of the Pearson correlation coefficients between three different pairs: $\langle\sigma_{1}^{z}\rangle$ and $\langle\sigma_{5}^{z}\rangle$, $\langle\sigma_{2}^{z}\rangle$ and $\langle\sigma_{4}^{z}\rangle$, as well as $\langle\sigma_{1}^{z}\rangle$ and $\langle\sigma_{2}^{z}\rangle$. We specifically consider two local magnetization spins to be synchronized when the Pearson coefficients $r_{ij}\approx 1$, which happens at times $\tau_{s}\approx4.5\pi$ for one-site noise, and $\tau_{s}\approx11.2\pi$ for two-site noise. When $\tau >\tau_{s}$, both $r_{15}$ and $r_{24}$ approach $1$, which means synchronous behavior emerge between the local magnetizations. In contrast, upon $r_{12}$ converges to $-1$, anti-synchronous behavior is established.

In order to extract the frequency of the oscillation, we perform the FFT on the time-dependent magnetizations. This method efficiently converts the data from the time domain to the frequency domain by computing the discrete Fourier transform of a time series \cite{DUHAMEL1990259}. The Fourier transform of the function $f(x)$ is $F(\omega)=\int_{-\infty}^{\infty}f(x)e^{-i\omega xdx}$.
\begin{figure}
\centering
\includegraphics[width=0.48\textwidth]{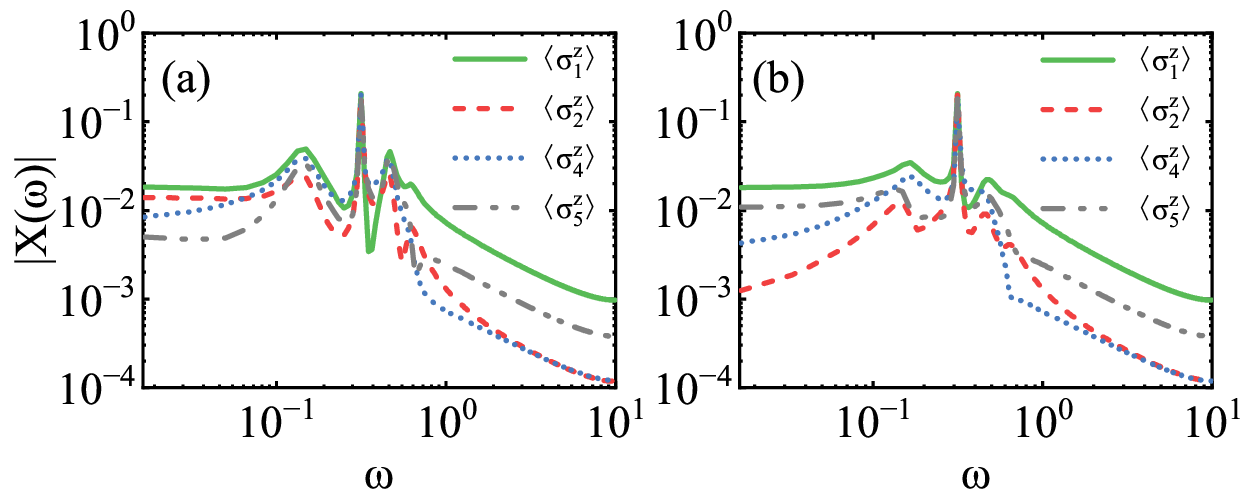}
\caption{The Fourier transform spectrum of the time dependent magnetizations $\sigma_{j}^{z}$ for two scenarios of noise acting on spin chain: one-site noise (as Fig. \ref{fig2}(b)) and two-site noise (as Fig. \ref{fig2}(c)).}
\label{fig4}
\end{figure}
In other words, it breaks down a signal into a combination of sine and cosine waves at various frequencies, which can be used to precisely reconstruct the original signal. Figure \ref{fig4} displays the results of FFT on the time-dependent magnetizations. The horizontal axis represents the frequency $\omega$, while the vertical axis represents the amplitude $|X(\omega)|$. We observe that each curve peaks around the same primary frequency $f$, approximately $0.31598$, and the amplitude corresponding to this frequency is consistent across all curves. It indicates that the system oscillates at a single frequency, with the expectation values of the local spin observables vibrating at the same amplitude after synchronization. This suggests that the system is completely synchronized (antisynchronized).
\begin{figure}[htbp]
\centering
\includegraphics[width=0.485\textwidth]{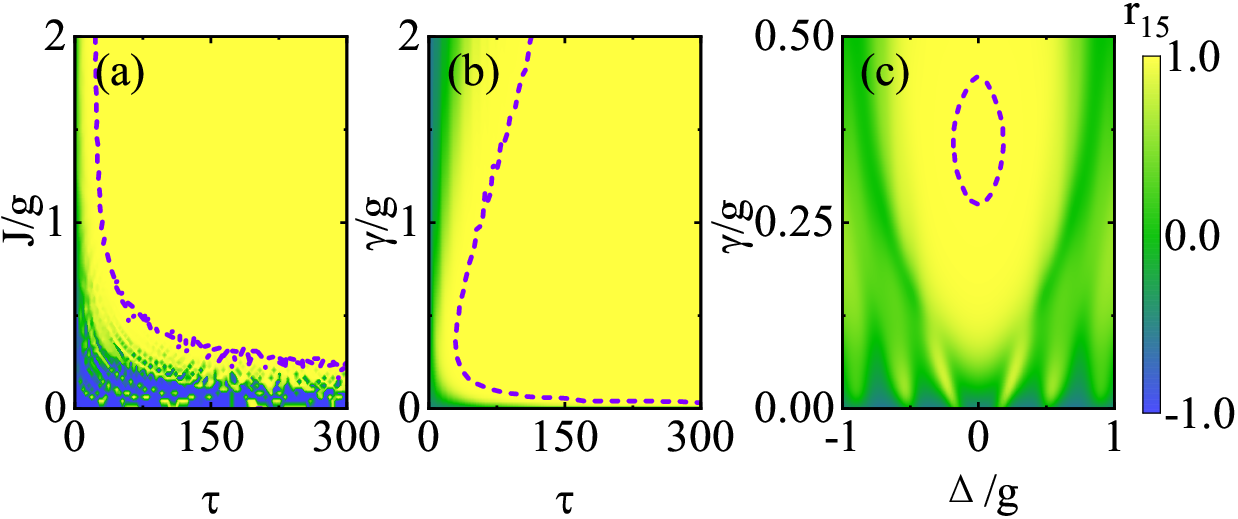}
\caption{(a), (b) The temporal evolution of the Pearson correlation coefficient $r_{15}$ between the local spin magnetizations $\sigma_{1}^{z}$ and $\sigma_{5}^{z}$, when change the nearest-neighbor interaction $J$ and the noise intensity $\gamma$. (c) The Pearson correlation coefficient $r_{15}$ as a functions of the noise amplitude $\gamma$ and detuning $\Delta$, exhibits a structure analogous to the Arnold tongue in classical synchronization theory. The purple dashed line corresponds to the curve $r_{15}=0.999$, signifying that these two local spin magnetizations have achieved synchronization. Other parameters are the same as in Fig. \ref{fig2}(b). }
\label{fig5}
\end{figure}

Considering a spin chain of length $N=6$ and focusing on the synchronization between spins $1$ and $5$ under the influence of a one-site noise $u=3$. We qualitatively discussed the impact of system parameters such as the interaction strength between spins $J$ and the noise intensity $\gamma$ on the synchronization time, as demonstrated in Fig. \ref{fig5}. Here, the purple dashed line represents $r_{15}\approx1$. The value of spin interactions affects the eigenfrequencies of $\langle\sigma_{j}^{z}\rangle$, thereby influencing the time to synchronization. However, the synchronization phenomenon does not vanish with changes in $J$. Increasing the noise amplitude correspondingly decreases the synchronization time, but there is an optimal amplitude. Beyond this amplitude, the synchronization time will increase, which is related to the quantum Zeno effect \cite{PhysRevA.109.062414}. To introduce a variable detuning $\triangle$ between the eigenfrequencies of the first and the fifth spins, we add the term
\begin{eqnarray}
\label{H1}
H_{1}=\frac{\Delta\hbar}{2}(\sigma_{1}^{z}-\sigma_{5}^{z}),
\end{eqnarray}
into the system's Hamiltonian Eq. (\ref{H}). Figure \ref{fig5}(c) displays the measured Pearson correlation coefficient $r_{15}$ at the onset of the synchronization regime $\tau_{s}\approx4.5\pi$ when both the noise amplitude $\gamma$ and the detuning $\triangle$ are varied. As expected from the general framework of synchronization, a structure resembling an Arnold tongue can be observed which defines the synchronized domain of classical synchronization phenomena \cite{BOCCALETTI20021,RevModPhys.77.137}. Specifically, the synchronization region is broadened as detuning is reduced and noise strength is increased.
\subsection{Loschmidt echo, purity, trace distance, quantum mutual information and entanglement of formation}
In parallel, we are also examining the influence of synchronization on the system and searching for quantum features of the synchronization. To explore the properties of the system's state, we assess the behaviors of the Loschmidt echo $L(\tau)$, purity and trace distance. We further examine this phenomenon by focusing on the informational aspects, namely the mutual information shared through quantum correlations among spins.

\begin{figure}
\centering
\includegraphics[width=0.48\textwidth]{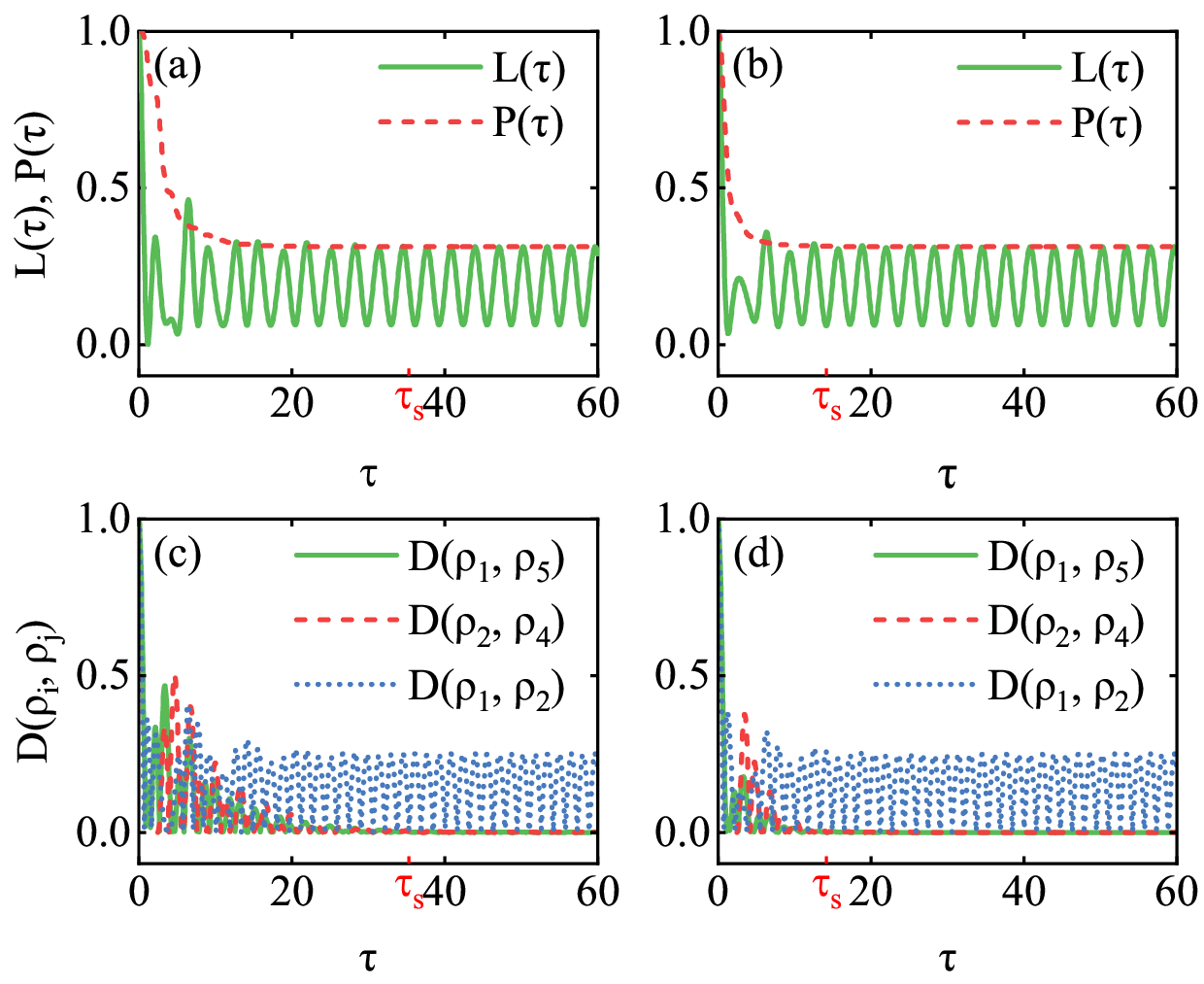}
\caption{(a), (b) The time evolution of the Loschmidt echo(green solid lines) and purity (red dashed lines). (c), (d) The dependence of the trace distances on $\tau$ between the local magnetization spins. (a), (c) Indicating the cases of one-site noise and other denote two-site noise case. The parameters are the same as in Figs. \ref{fig2}(b) and \ref{fig2}(c).}
\label{fig6}
\end{figure}

The Loschmidt echo is based on the concept of capturing subtle changes in the system's state after it is perturbed. It has been used to probe for quantum metrology \cite{PhysRevA.94.010102}, quantum phase transition \cite{PhysRevLett.96.140604,PhysRevB.89.125120} and quantum chaos \cite{PhysRevLett.124.160603}. The Loschmidt echo is defined as \cite{PhysRevLett.91.210403}
\begin{eqnarray}
\label{Loschmidt echo}
L(\tau)=Tr[\rho^{\dag}(\tau)\rho(0)],
\end{eqnarray}
which quantifies the overlap between the initial state $\rho(0)$ and the state $\rho(\tau)$ at time $\tau$, thereby characterizing the decoherence of the system \cite{PhysRevA.75.032333}. In Fig. \ref{fig6}, we present the time dependence of the Loschmidt echo $L(\tau)$. At time $\tau=0$, the system is in its initial state, with the Loschmidt echo value being $1$. The interference from noise causes the system to undergo decoherence, leading to a sharp decline in the amplitude of the Loschmidt echo within a short time. However, the occurrence of stable synchronization leads to the existence of a decoherence-free subspace \cite{PhysRevLett.132.010402,PhysRevLett.132.196601}, in which quantum information is protected against the effects of decoherence \cite{Lidar2003}. The Loschmidt echo oscillates steadily over time, reflecting stable quantum phase correlations during the evolution of the quantum state \cite{PhysRevLett.118.015701}.

The purity is used to describe the degree of mixedness of a quantum state and is defined as
\begin{eqnarray}
\label{purity}
P(\tau)=Tr[\rho^{2}(\tau)].
\end{eqnarray}
We show how the system's purity evolves over time in Fig. \ref{fig6}. Under the influence of noise, the initial pure state of the system collapses into a mixed state. Unlike the Loschmidt echo, the purity remains constant, once the stable oscillation of the magnetization is established \cite{PhysRevA.107.032219}.

We further analyse the relationship between the states of three pairs of spins: $\rho_{1}(\tau)$ and $\rho_{5}(\tau)$, $\rho_{2}(\tau)$ and $\rho_{4}(\tau)$, as well as $\rho_{1}(\tau)$ and $\rho_{2}(\tau)$, during the time evolution. $\rho_{j}(\tau)=Tr_{\neq j}[\rho(\tau)]$ is the reduced density matrix of the $jth$ spin which is obtained by taking the partial trace of the density matrix of the total system. The trace distance \cite{PhysRevA.93.012110,Aaronson_2013} characterizes the similarity between two reduced density matrices, $\rho_{i}$ and $\rho_{j} $, is given by
\begin{eqnarray}
\label{trace distance}
D(\rho_{i},\rho_{j})=\frac{1}{2}Tr\left[\sqrt{(\rho_{i}-\rho_{j})^{\dag}(\rho_{i}-\rho_{j})}\right],
\end{eqnarray}
where $0\leq D(\rho_{i},\rho_{j})\leq 1$ for two states. Figure \ref{fig6}(c) and (d) show the time evolution of the trace distances between the local magnetization spins. Initially, the states were fully distinguishable, and then their degree of distinguishability gradually decreases. After synchronization, the trace distance between two states of completely synchronized spins (the local magnetization spin of the sites $1$ and $5$, as well as $2$ and $4$) gradually becomes zero over time. These two quantum states are mathematically identical and indistinguishable by all possible measurement, which is crucial for achieving high-fidelity quantum state transfer on spin chains \cite{RevModPhys.88.041001,PhysRevA.88.032303}. For fully antisynchronized spins (the local magnetization spin of the sites $1$ and $2$), the trace distance between the corresponding reduced density matrices exhibits periodic oscillations which is consistent with the evolution of $\langle\sigma_{j}^{z}\rangle$ over time.
\begin{figure}
\centering
\includegraphics[width=0.48\textwidth]{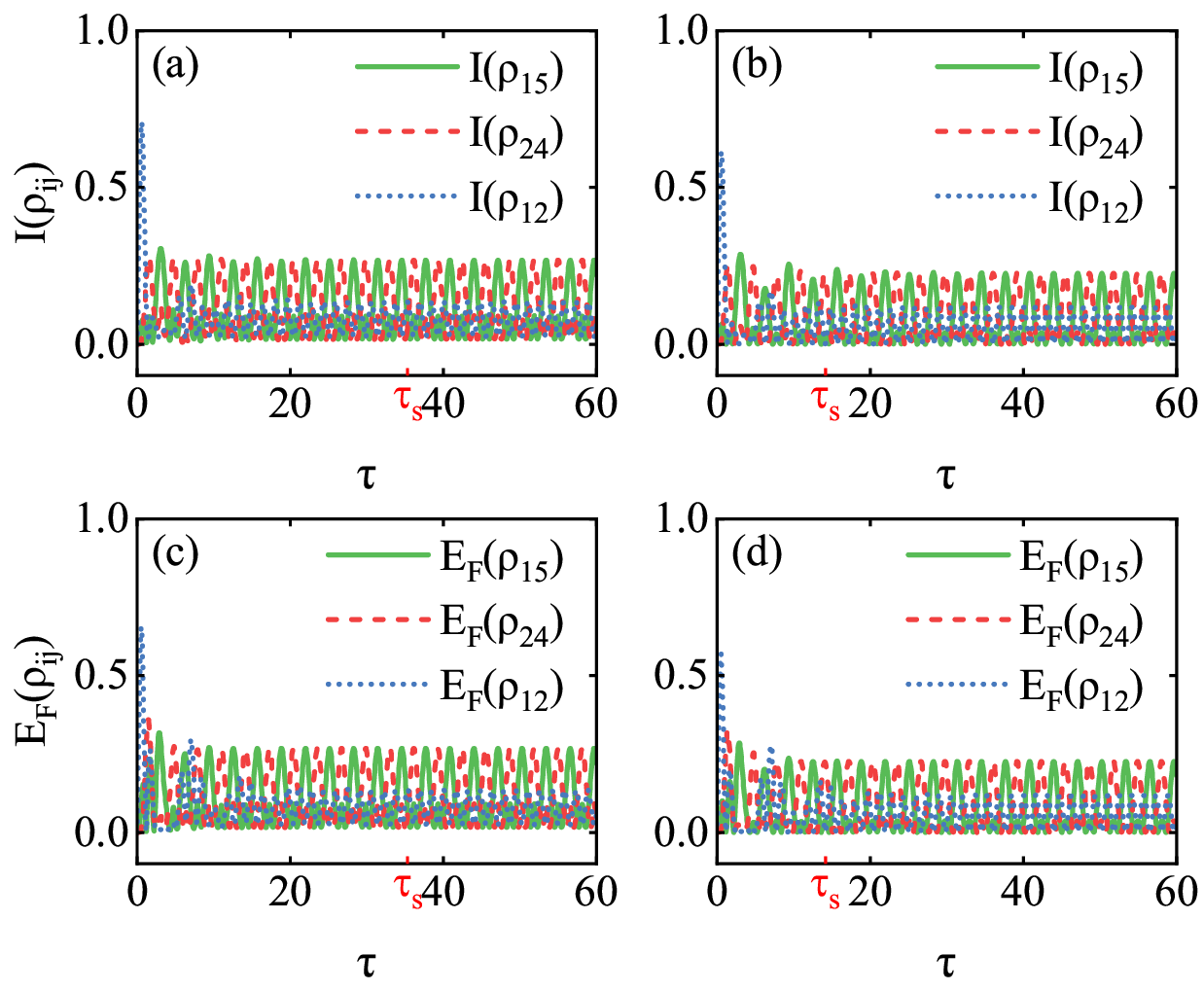}
\caption{Mutual information and the  entanglement of formation as a function of time $\tau$ between the local magnetization spins. One-site noise is represented in cases (a), (c) whereas two-sites noise is depicted in cases (b), (d). The parameters are the same as in Figs. \ref{fig2}(b) and \ref{fig2}(c).}
\label{fig7}
\end{figure}

In quantum systems, noise typically induces decoherence, which degrades the coherence and entanglement of quantum states, thereby diminishing the quantum characteristics of the system \cite{PhysRevB.109.165306,PhysRevB.67.094510,PhysRevA.94.042110}. In this regard, we discuss the correlations between spins by examining mutual information and entanglement formation within the quantum system. Here, we quantify the total amount of classical and quantum correlations in the bipartite system state $\rho_{i},\rho_{j}$ using quantum mutual information \cite{PhysRevA.91.012301,RevModPhys.80.517}.
\begin{eqnarray}
\label{MI}
I(\rho_{ij})=S(\rho_{i})+S(\rho_{j})-S(\rho_{ij}),
\end{eqnarray}
where $\rho_{j}$ is the reduced density matrices of subsystem, $\rho_{ij}(\tau) = Tr_{\{1, \ldots, n\} \setminus \{i,j\}}[\rho(\tau)]$ and $S(\rho_{j})=Tr(\rho_{j}log\rho_{j})$ is its von Neumann entropy.

Various tools have been proposed for the degree of quantumness of the state of a bipartite system. Among them, the entanglement of formation \cite{RevModPhys.80.517,RevModPhys.81.865} $E_{F}$ is a well established measure that quantifies the number of singlet states that are necessary to prepare a given entangled state using only local operations and classical communication. The entanglement of formation $E_{F}$ is defined as
\begin{eqnarray}
\label{EF}
E_{F}(\rho_{ij})=H_{b}\left(\frac{1-\sqrt{1-E_{C}^{2}(\rho_{ij})}}{2}\right),
\end{eqnarray}
where $H_{b}$ is the binary entropy, $H_{b}(q)=:-qlog_{2}q-(1-q)log_{2}(1-q)$. The concurrence $E_{C}(\rho_{ij})= max(0,\mu_{1}-\mu_{2} -\mu_{3}-\mu_{4})$, where $\mu_{i}$ are the eigenvalues of the Hermitian matrix $R =\sqrt{\sqrt{\rho_{ij}}\tilde{\rho}_{ij}\sqrt{\rho_{ij}}}$ in decreasing order $\tilde{\rho}_{ij}$ is the spin flipped matrix of $\rho_{ij}$.

Noise is commonly considered detrimental to quantum features due to decoherence. When synchronization cannot be established under the influence of noise, the entanglement of formation between subsystems becomes negligible. A sufficient condition for stable synchronization is the existence of a decoherence-free subspace with only one single eigenmode \cite{PhysRevLett.132.010402}, which allows the system to sustain a non-decaying mode immune to external perturbations. As shown in Fig. \ref{fig7}, after brief transient dynamics, both the mutual information and the entanglement of formation between synchronized and antisynchronized spins exhibit steady oscillations with nonzero amplitude. Therefore, the total correlations (including both classical and quantum components) and the entanglement of formation between the synchronized or antisynchronized spins remain relatively stable. In these correlations, entanglement constitutes the majority. Moreover, it is observed that for spins in anti-synchronization, the trace distance is smaller when the mutual information and entanglement between the two subsystems are stronger, and conversely, the trace distance becomes larger when the mutual information and entanglement are weaker.
\section{Conclusions} \label{section4}
We have investigated the synchronization behavior of a quantum spin chain with periodic boundary conditions under the influence of local Gaussian white noise. The presence of noise induces a separation of timescales in the decay rates of the system's eigenmodes, thereby governing the system's evolution and enabling the observation of synchronization and antisynchronization in local spin observables. Stable synchronization conditions for one-site noise and two-site noise have been derived through perturbation theory within the Liouville space. FFT has been performed on the time-dependent magnetizations, thereby obtaining the oscillation frequencies. We have also found that the system evolves into a stable mixed state after synchronization while the Loschmidt echoes exhibit stable oscillatory behavior which plays a significant role in experimentally detecting oscillatory dynamics. Further exploration of the mutual information and entanglement between synchronized and antisynchronized spins demonstrates that these spins exhibit quantum correlations after synchronization. Quantum spin chains can serve as quantum wires for connecting quantum devices \cite{SUR2020126176} and as possible quantum channels for quantum state transfer and entanglement dynamics \cite{PhysRevLett.91.207901,PhysRevA.104.012410}. Due to the quantum synchronization induced by noise in spin chains exhibiting robustness against perturbations, our study could potentially be applied to quantum communication and quantum computing based on synchronization \cite{doi:10.1142/S0219749911008180}.
\section*{Acknowledgments}
The work is supported by the National Natural Science Foundation of China (Grant No. 12475026) and the Natural Science Foundation of Gansu Province (No. 25JRRA799).
\appendix
\section{The calculation of the decay rates} \label{appendix1}
In this appendix, we provide the detailed procedure for calculating the decay rates $m_{kl}^{u}$. For the case of non-degenerate eigenfrequencies, we need to compute the expectation of the perturbation
\begin{flalign}
\label{mklnon}
m_{kl}^{u}|_{\text{non}} &=2\langle\!\langle\varphi_{k},\varphi_{l}|\mathcal{Y}|\varphi_{k},\varphi_{l}\rangle\!\rangle &\nonumber \\
&=2\langle\!\langle\varphi_{k},\varphi_{l}|[Y\otimes \mathbb{I}-\mathbb{I}\otimes Y]^{2}|\varphi_{k},\varphi_{l}\rangle\!\rangle   &\nonumber\\
&=2\langle\!\langle\varphi_{k},\varphi_{l}| Y^{2}\otimes\mathbb{I}+\mathbb{I}\otimes Y^{2}-2Y\otimes Y|\varphi_{k},\varphi_{l}\rangle\!\rangle  &\nonumber\\ &=\frac{4}{N}\left(\sin(\frac{uk\pi}{N})^{2}+\sin(\frac{ul\pi}{N})^{2}\right) & \nonumber\\
&-\frac{16}{N^{2}}\left(\sin(\frac{uk\pi}{N})^{2}\sin(\frac{ul\pi}{N})^{2}\right).
\end{flalign}
For every twofold degenerate frequencies, there are two orthonormal eigenvectors $|a\rangle\!\rangle=|\varphi_{k},\varphi_{l}\rangle\!\rangle,
|b\rangle\!\rangle=|\varphi_{N-k},\varphi_{N-l}\rangle\!\rangle$.
We have to compute the eigenvalues and eigenvectors of the perturbation matrix
\begin{eqnarray}
\label{perturbation matrix}
P=\begin{pmatrix}2(\mathcal{Y}^2)_{aa}&2(\mathcal{Y}^2)_{ab}\\\\2(\mathcal{Y}^2)_{ba}&2(\mathcal{Y}^2)_{bb}\end{pmatrix},
\end{eqnarray}
with $(\mathcal{Y}^2)_{ab}= \langle\!\langle a|(\mathcal{Y}^2)|b\rangle\!\rangle$. Because $\Omega$ and $Y$ are real and symmetric,
we have $(\mathcal{Y}^2)_{ab}=(\mathcal{Y}^2)_{ba}$.
\begin{flalign}
\label{sin}
\sin(\frac{u(N-k)\pi}{N})=&\sin(u\pi-\frac{uk\pi}{N})\nonumber \\=&(-1)^{u+1}\sin(\frac{uk\pi}{N}). &
\end{flalign}
Thus,
\begin{flalign}
\label{pm value}
(\mathcal{Y}^2)_{aa}=(\mathcal{Y}^2)_{bb}=&\sin(\frac{uk\pi}{N})^{2}+\sin(\frac{ul\pi}{N})^{2}&\nonumber\\
&-2\sin(\frac{uk\pi}{N})^{2}\sin(\frac{ul\pi}{N})^{2}, &\\
(\mathcal{Y}^2)_{ab} =(\mathcal{Y}^2)_{ba}=&-2\sin(\frac{uk\pi}{N})\cos(\frac{ul\pi}{N}) \nonumber\\
&\sin(\frac{ul\pi}{N})\cos(\frac{uk\pi}{N}). &
\end{flalign}
The decay rate and zeroth order eigenvector thus simplify to
\begin{flalign}
\label{mkldeg}
m_{kl}^{u}|_{deg} &=2(\mathcal{Y}^2)_{aa}\pm 2|(\mathcal{Y}^2)_{ab}|,
\end{flalign}
\begin{eqnarray}
\label{vectors}
v^{+}=\left(\frac{(\mathcal{Y}^2)_{aa}}{|(\mathcal{Y}^2)_{ab}|},1\right)^{T}, v^{-}=\left(-\frac{(\mathcal{Y}^2)_{aa}}{|(\mathcal{Y}^2)_{ab}|},1\right)^{T}.
\end{eqnarray}
From Eq. (\ref{vectors}) the perturbed eigenvectors can be directly identified as
\begin{eqnarray}
\label{perturbed eigenvectors}
|a\rangle\!\rangle=\frac{1}{\sqrt{2}}\left(1\pm \text{sgn}\{(\mathcal{Y}^2)_{ab}\}|a\rangle\!\rangle\right).
\end{eqnarray}
The projection onto the magnetization eigenbasis produces the same magnetization eigenmode for $|a\rangle\!\rangle $ and $|b\rangle\!\rangle$. That is $\langle\!\langle\sigma_{j}^{z}|a\rangle\!\rangle=\langle\!\langle\sigma_{j}^{z}|b\rangle\!\rangle$. We therefore finally obtain for the zeroth-order eigenmode
\begin{eqnarray}
\label{zeroth-order eigenmode}
\langle\!\langle\sigma_{j}^{z}|a\rangle\!\rangle=\frac{1}{\sqrt{2}}\left(1\pm\text{sgn}\{(\mathcal{Y}^2)_{ab}\}\right)\langle\!\langle\sigma_{j}^{z}|a\rangle\!\rangle.
\end{eqnarray}
Depending on the sign of $(\mathcal{Y}^2)_{ab}$ one of the corrections will always vanish while the other gains a factor of $\sqrt{2}$. The decay rates are thus given by
\begin{flalign}
\label{mklnon}
m_{kl}^{u}|_{\text{deg}} &=2(\mathcal{Y}^2)_{aa}- 2|(\mathcal{Y}^2)_{ab}|  &\nonumber \\ &=\frac{4}{N}\left(\sin(\frac{uk\pi}{N})^{2}+\sin(\frac{ul\pi}{N})^{2}\right) & \nonumber\\
&-\frac{32}{N^{2}}\left(\sin(\frac{uk\pi}{N})^{2}\sin(\frac{ul\pi}{N})^{2}\right).
\end{flalign}

When considering two-site noise with $V =\sigma_{u}^{z}+\sigma_{v}^{z}$, a comparable approach can be applied. We can obtain the decay rates
\begin{equation}
\begin{aligned}
\label{two site nondegenerate}
m_{kl}^{u,v}|_{\text{non}} ={}&\frac{4}{N}\left(\sin(\frac{uk\pi}{N})^2+\sin(\frac{vk\pi}{N})^2\right. \\ &+\sin(\frac{ul\pi}{N})^2+\sin(\frac{vl\pi}{N})^2\Biggr) \\&-\frac{16}{N^2}\left(\sin(\frac{uk\pi}{N})^2+\sin(\frac{vk\pi}{N})^2\right)\times \\
&\quad\left(\sin(\frac{ul\pi}{N})^2 + \sin(\frac{vl\pi}{N})^2\right),
\end{aligned}
\end{equation}
for nondegenerate eigenfrequencies, and
\begin{equation}
\begin{aligned}
\label{two site degenerate}
m_{kl}^{u,v}|_{\text{deg}} ={}&m_{k l}^{u,v}|_{\text{non}}-\frac{16}{N^2}\left(\sin(\frac{uk\pi}{N})\sin(\frac{ul\pi}{N})\right. \\
&\left.+\sin(\frac{vk\pi}{N})\sin(\frac{vl\pi}{N})\right)^2,
\end{aligned}
\end{equation}
for degenerate eigenfrequencies.

\bibliography{refercence}
\end{document}